\documentclass[prb,aps,twocolumn,showpacs]{revtex4}

\usepackage{amsmath,amsthm,amssymb}

\usepackage{amsmath,graphicx,bm}
\usepackage{color,ulem}
\newlength{\figwidth}
\newlength{\figlarge}
\begin{document}
\title{Topological phase in a two-dimensional metallic heavy-fermion system}

\author{Tsuneya Yoshida} 
\author{Robert Peters} 
\author{Satoshi Fujimoto} 
\author{Norio Kawakami}

\affiliation{Department of Physics, Kyoto University, Kyoto 606-8502, Japan}

\date{\today}
\begin{abstract}
We report on a topological insulating state in a heavy-fermion system away from half-filling, which is hidden within a ferromagnetic metallic phase.
In this phase, the cooperation of the RKKY interaction and the Kondo effect, together with the spin-orbit coupling, induces a spin-selective gap, bringing about topologically non-trivial properties. This topological phase is robust against a change in the chemical potential in a much wider range than the gap size. We analyze these remarkable properties by using dynamical mean field theory and the numerical renormalization group. Its topological properties support a gapless chiral edge mode, which exhibits a non-Tomonaga-Luttinger liquid behavior due to the coupling with bulk ferromagnetic spin fluctuations. We also propose that the effects of the spin fluctuations on the edge mode can be detected via the NMR relaxation time measurement.
\end{abstract}
\maketitle

\section{Introduction}
In this decade, $\mathbb{Z}_2$ topological band insulators (TBIs) have been found as new classes of quantum phases.\cite{Hasan10,Qi10} Their topological order is described in terms of the corresponding topological invariant and reflected in the existence of gapless edge modes which are protected by time-reversal symmetry. Such phases have been experimentally found in $\mathrm{HgTe}/\mathrm{CdTe}$ quantum wells \cite{exp_2D-QW_MKonig} and bismuth based compounds.\cite{exp_3D-bismuth_DHsieh,exp_3D-bismuth_YXia,exp_3D-bismuth_YLChen}
These discoveries have stimulated the study of topologically non-trivial phases, providing a new research arena in condensed matter physics. One of the current major issues is the influence of strong correlations upon these topological phases, because electron correlations under such non-trivial conditions are expected to spark new and exotic phenomena. This issue has further been stimulated by theoretical proposals for the realization of topological phases in $d$- and $f$- electron systems,\cite{NaIrO_Nagaosa09,TMI_LBalents09,Heusler_Chadov10,Heusler_Lin10,skutterudites_Yan12} triggering theoretical studies on the competition between topological phases and long-range ordered phases\cite{Yamaji11,AFvsTBI_DQMC_Hohenadler11,AFvsTBI_DQMC_edge_early_Zheng10,AFvsTBI_VCA_Yu11,AFvsTBI_CDMFTWu11,Rachel10,Varney10,Wang10,TBI_Mott_Yoshida,TBI_Mott_Tada} and topologically non-trivial phases induced by the Coulomb interaction.\cite{Raghu08,TM-Z_YZhang09,Kurita11} 

Heavy-fermion systems, which are representative examples for strongly correlated systems, have exhibited various intriguing properties due to the competition of the RKKY interaction and the Kondo effect, such as huge enhancement of the effective mass, metamagnetism, orbital order, unconventional superconductivity near the critical point, etc. A recent remarkable proposal in  heavy-fermion systems is a {\it topological Kondo insulator}, first, discussed by Dzero \textit{et
  al.}.\cite{TKI_Dzero10} Its properties have been further studied in ref. \cite{TKI_Dzero12,TKI_Tran12}, and the first-principle calculations have predicted that the filled skutterudites $\mathrm{CeOs_4As_{12}}$ \cite{skutterudites_Yan12} and $\mathrm{SmB_6}$ \cite{TKI_SmB6_Takimoto,TKI_SmB6_Lu,TKI_SmB6_Botimer,TKI_SmB6_Zhang} can be strong topological Kondo insulators.  

All the above topological insulators have been proposed, of course, for the insulating phases. We now ask the question: {\it Is it possible to realize a topologically non-trivial phase within a metallic phase away from a commensurate filling}? Naively, it seems to be impossible to find a topological insulator, if the noninteracting system is metallic. However, we will demonstrate that it is possible to find a topological insulating state even within a metallic phase, if strong electron correlations are present.

In this paper, we propose a spin-selective topological insulator in a two-dimensional heavy-fermion system, which is hidden within a ferromagnetic metallic phase. For the stabilization of this phase, the RKKY interaction and the Kondo effect, together with the spin-orbit coupling, play the principal role. Furthermore, these effects support the robustness of this topological phase against shifts of the chemical potential, which is  much larger than the gap size. The topologically non-trivial structure gives rise to a chiral gapless mode at the boundaries. With a phenomenological treatment, we clarify that this gapless mode shows a non-Tomonaga-Luttinger behavior due to the coupling to bulk magnetic fluctuations. 

\section{Model and method}
To study the topological properties of two-dimensional heavy-fermion systems, we employ the periodic Anderson model with the spin-orbit coupling, which is an extension of the Bernevig-Hughes-Zhang model.\cite{Bernevig06_BHZ} The corresponding Hamiltonian reads 
\begin{widetext}
\begin{eqnarray}
  H&=& \sum_{i,\alpha,\sigma} \varepsilon_\alpha n_{i,\alpha,\sigma} -\sum_{ i,j ,\sigma} \hat{c}^{\dagger}_{i,\alpha,\sigma} \hat{t}_{i,j,\alpha,\alpha',\sigma} \hat{c}_{j,\alpha',\sigma} + U\sum_{i} n_{i,f,\uparrow} n_{i,f,\downarrow},\\
-\hat{t}_{i,j \sigma} &=&\left(
\begin{array}{cc}
 -t_{ff}(\delta_{i,j\pm \hat{x}}+\delta_{i,j\pm \hat{y}})& t_{cf}(-\mathrm{sign}(\sigma)(\delta_{i,j+\hat{y}}-\delta_{i,j-\hat{y}})+i(\delta_{i,j+\hat{x}}-\delta_{i,j-\hat{x}}))  \nonumber \\
t_{cf}(\mathrm{sign}(\sigma)(\delta_{i,j+\hat{y}}-\delta_{i,j-\hat{y}})+i(\delta_{i,j+\hat{x}}-\delta_{i,j-\hat{x}})) & t_{cc}(\delta_{i,j\pm \hat{x}}+\delta_{i,j\pm \hat{y}})
\end{array}
\right),
\label{hamiltonian}
\end{eqnarray}
\end{widetext}
where $n_{i,\alpha,\sigma}=c^{\dagger}_{i,\alpha,\sigma}c_{i,\alpha,\sigma}$. The operator $c^{\dagger}_{i,\alpha,\sigma}(c_{i,\alpha,\sigma})$ creates (annihilates) an electron at site $i$ and in orbital $\alpha=c,f$ and spin $\sigma=\uparrow, \downarrow$ state. The hopping integral $t_{ff}$ is assumed to be much smaller than $t_{cc}$ in order to describe a heavy-fermion system. The hybridization term ($t_{cf}$) between conduction and $f$-electrons has the specific form characteristic of the spin-orbit coupling,\cite{Bernevig06_BHZ} and can drive the system into a topologically non-trivial band insulator.
{
We analyze the system by using dynamical mean field theory (DMFT),\cite{DMFT_Georges,DMFT_Metzner,DMFT_Muller-Hartmann} which treats local correlations exactly by mapping the original lattice model onto an effective impurity model. We employ the numerical renormalization group (NRG) method to solve the impurity model.\cite{NRG_Wilson,NRG_Bulla} In this study we set the model parameters as $t_{ff}=0.2t_{cc} $, $t_{cf}=t_{cc}$, $\epsilon_{c} =0$, and $\epsilon_{f} =-8t_{cc}$. The hopping integral $t_{cc}$ is chosen as the energy unit.

\section{Bulk Properties}
Let us start with the half-filled case,\cite{TKI_Tran12} where each lattice site is occupied by one conduction electron in average ($N_{tot}=2$). The DMFT results are summarized in the phase diagram (Fig. \ref{fig:phase}). At half-filling, we can observe the competition between the RKKY interaction and the Kondo effect. Namely, in the strong coupling region ($U>12$), the RKKY interaction is dominant and the system is an antiferromagnetic insulator (AFI). When the interaction is decreased, the Kondo effect becomes dominant, and the system enters the Kondo insulating phase for $U< 12$, where the Kondo gap is formed (i.e., the system is in a paramagnetic phase.). Note that for $4.8 < U < 12$, this Kondo insulator has non-trivial topological properties, which can be detected via the spin Chern number. 
This paramagnetic "topological Kondo insulator" is essentially identical to those proposed previously,\cite{TKI_Dzero10,TKI_Dzero12,TKI_Tran12} although the models studied there are slightly different from the present one. 
At half-filling, the competition between the RKKY interaction and the Kondo effect plays an essential role in the above-mentioned behaviors. On the other hand, away from half-filling, the above two effects cooperate with each other and stabilize a novel phase. In the following, we discuss the hole-doped case.
\begin{figure}[!h]
\begin{center}
\includegraphics[width=85mm,clip]{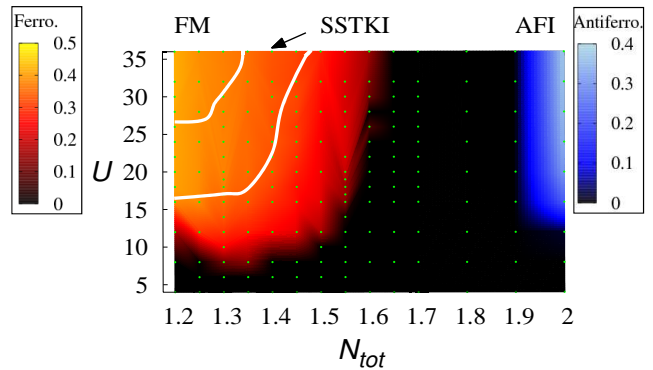}
\end{center}
\caption{(Color Online).
Zero-temperature phase diagram as functions of interaction strength $U$ and total filling $N_{tot}$, obtained by the DMFT+NRG method. The colors represent the magnetic moments of the $f$-electrons for each phase; for the ferromagnetic (red) (antiferromagnetic (blue)) phase, respectively. 
At half-filling, the system is an antiferromagnetic insulator (AFI). For $N_{tot}\lesssim 1.6$, a ferromagnetic metallic (FM) phase is realized. In the region enclosed by two white lines, a spin-selective topological Kondo insulator (SSTKI) is realized; the majority-spin state forms a bulk gap which induces non-trivial properties.
}
\label{fig:phase}
\end{figure}

The antiferromagnetic order vanishes when the system is doped, and it enters a paramagnetic metallic phase for $1.6 \lesssim N_{tot} \lesssim 1.9$, as shown in Fig. \ref{fig:phase}.  Upon further doping, the system shows a ferromagnetic order for $N_{tot} \lesssim 1.6 $. These properties can be naturally expected for the ordinary Kondo lattice systems. A remarkable point is that a "topological insulator" is hidden within this ferromagnetic metallic phase. To confirm the existence of the gap, we plot the spin-dependent local density of states (LDOS) for several values of interaction strength in Fig. \ref{fig:total_DOS}. In this figure, we assume that the majority-spin electrons are up-spin electrons ($N_{\uparrow}>N_{\downarrow}$, where $N_{\sigma}=\langle n_{c,\sigma} +n_{f,\sigma}\rangle$). 
At $U=12$, where the system is in the Kondo regime, the $f$-electrons are weakly magnetized, and no gap is observed, as seen in Fig. \ref{fig:total_DOS}. 
On the other hand, at $U=24$, the RKKY interaction enhances the spin-polarization and induces a spin-selective gap for the majority-spin electrons (i.e., for the majority-spin state, a metal to insulator transition is induced). This gap results from cooperation of the RKKY interaction and the Kondo effect, and the resulting gap size is indeed of the order of the Kondo temperature. The mechanism of forming the spin-selective gap is identical to that proposed for the topologically trivial heavy-fermion systems.\cite{Ferro_KI_Beach08,Ferro_KI_Peters11,Ferro_KI_Howczak,Ferro_KI_Peters12}
Namely, in the strong correlation regime where the $f$-electron is localized, the gap formation is due to the formation of a local Kondo singlet which is composed of all conduction electron in the up-spin state ($c,\uparrow$) and all $f$-electrons in the down-spin state ($f,\downarrow$) as well as a part of conduction electrons in the down-spin state ($c,\downarrow$) and $f$-electrons in the up-spin state ($f,\uparrow$). The remaining parts of electrons in state ($f,\uparrow$) and ($c,\downarrow$) contribute to the formation of ferromagnetism. The essence of the spin-selective gap formation is that this partial Kondo screening leads to a commensurability condition for the majority-spin (up-spin) electrons, which in turn energetically stabilizes this half-metallic state.\cite{Ferro_KI_Peters11} The spin-selective behavior still holds even if the localized $f$-electrons are itinerant. 
Because the system is stabilized via the formation of the Kondo gap, once the commensurability condition is satisfied, the system tends to form the spin-selective gap. The cooperation of these two effects, the RKKY interaction and the Kondo screening, supports the robustness against a shift of the chemical potential, as discussed below in more detail. However, if the interaction becomes extremely strong, the Kondo effect itself is suppressed, and the system favors the spontaneous magnetization rather than the Kondo-singlet formation. Hence the majority-spin states have finite density of states at the Fermi energy, making the system metallic. 
This behavior can be observed in Fig. \ref{fig:total_DOS} for $U=36$, and is also observed in the ordinary Kondo lattice model.\cite{Ferro_KI_Zitko} We note that the formation of the spin-selective gap is expected for any dimensions, and indeed is confirmed in one- \cite{Ferro_KI_Peters12} and infinite-dimensional \cite{Ferro_KI_Peters11} Kondo lattice model.
The remarkable feature of the spin-selective insulating phase, studied in this paper, is its topological structure, which emerges away from half-filling, where one usually does not expect any topological phase.
The topological structure can be detected via the Chern number for the majority-spin state. 
The corresponding topological invariant is described in terms of the Green's function as follows:\cite{Volovik03,Gurarie11}
\begin{figure}[!h]
\begin{center}
\includegraphics[width=90mm,clip]{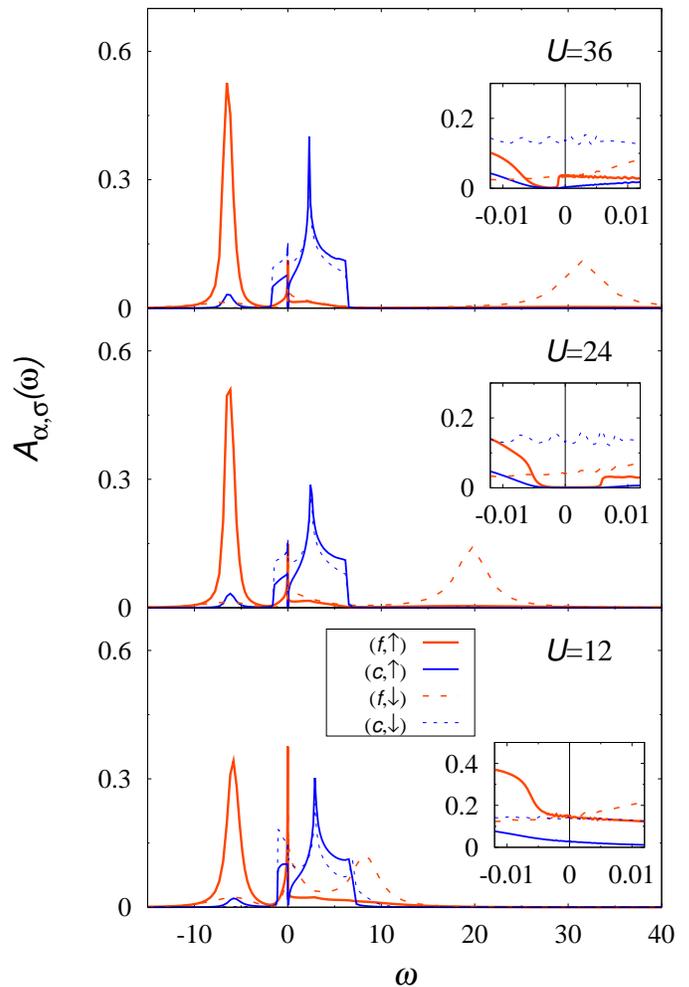}
\end{center}
\caption{(Color Online). Spectral functions ($A_{\alpha,\sigma}=-\mathrm{Im}G^R_{\alpha, \alpha,\sigma}(\omega)/\pi$) at $N_{tot}=1.3$ for several values of interaction strength. Inset: the spectral functions around the Fermi energy $\omega=0$ (denoted by the vertical line).
For this set of parameters, the system is in the ferromagnetic phase. Besides, at $U=24$, the up-spin state has a bulk gap. 
With increasing or decreasing interaction strength, a metal-insulator transition occurs for the majority-spin states due to the shift of the bulk gap.
}
\label{fig:total_DOS}
\end{figure}
\begin{eqnarray}
N_{\sigma} &=& \frac{\epsilon^{\mu \nu \rho}}{24\pi^2} \int d^3p \; \mathrm{tr}[ \hat{G}_{\sigma}(p)^{-1} \frac{\partial \hat{G}_{\sigma}(p)}{\partial p_{\mu}}  \hat{G}_{\sigma}(p)^{-1} \nonumber \\ 
           &&\frac{\partial \hat{G}_{\sigma}(p)}{\partial p_{\nu}} \hat{G}_{\sigma}(p)^{-1} \frac{\partial \hat{G}_{\sigma}(p)}{\partial p_{\rho}}   ]. 
\end{eqnarray}
Here, the notation $p=(i\omega,\textrm{\boldmath $p$})$ is used. As long as the Green's function does not have singularity, this quantity can be simplified as \cite{Wang12} 
\begin{eqnarray}
N^{ch}_{\sigma} &=& \frac{1}{4\pi} \int d^2k \hat{d}(k) \cdot \partial_{k_x}\hat{d}(k) \times  \partial_{k_y}\hat{d}(k) \\
\hat{h'}_{\sigma}(k) &=& \bm{I}d_{0,\sigma}(k) +\sum_{i}d_{i,\sigma}(k)\hat{\sigma}_i \label{eq: h_and_d},
\end{eqnarray}
{where $\hat{h'}_{\sigma}(k)=\hat{h}_{\sigma}(k)-\bm{I}\mu+\hat{\Sigma}_{\sigma}(i\omega=0)$ ($\hat{h}(k)$, $\hat{\sigma}_i$ and $\bm{I}$ are the Fourier transform of the hopping matrix, Pauli matrix for the orbital space and the identity matrix, respectively). $d_{i}$ is defined by Eq. (\ref{eq: h_and_d}) and, $\hat{d}_i=d_i/|d|$. The results are plotted in Fig. \ref{fig:ch} (a). 
As seen in this figure, the up-spin (majority) electrons possess the non-trivial structure, $N^{ch}_{\uparrow}=1$. We thus end up with a spin-selective topological insulator, which is embedded in a ferromagnetic metallic phase in a low-density region of our heavy-fermion system. 
Here, we note again that for $U>30$ or $U<18$, the majority-spin states have the finite density of states at the Fermi energy and the topological invariant is no longer well-defined.
This topological structure is reflected in a quantized Hall conductivity for the majority-spin electrons. In experiments, however, a finite but non-quantized spin-Hall conductivity may be detected due to the contribution of the gapless bulk minority-spin electrons.  Although the non-trivial phase is realized for the above choice of the parameters, one might wonder whether it is stabilized for other choices of the parameters. 

As seen in the inset of Fig. \ref{fig:total_DOS}, the spin-selective gap is $\Delta \sim 0.01$. This implies that the gap size (i.e. Kondo temperature) is approximately 100K provided that $t_{cc} \sim 1$eV.  For typical heavy-fermion compounds, the gap size is tens of Kelvin (e.g., for $\mathrm{SmB_6}$, which is a well-known Kondo insulator, the gap is approximately 40K\cite{TKI_SmB6_Botimer}). Thus, the size of the spin-selective gap in our model is realistic. However, one may still wonder how the change in the electron filling affects this topological phase; this change causes a shift in the chemical potential to a higher or lower energy region and might immediately spoil the topological structure in the ferromagnetic phase. 
Contrary to this naive speculation, the above spin-selective gap is rather robust against such changes in the chemical potential because of the correlation effects; the spin-selective gap is caused by the interplay of the RKKY interaction and the Kondo effect.\cite{Ferro_KI_Peters11} Namely, even if the chemical potential is shifted to outside of the spin-selective gap, these two effects can reconstruct the electronic structure to satisfy the commensurability condition and thus restore the spin-selective gap at the Fermi energy. Hence, although the change in the electron filling induces the shift of the chemical potential, the position of the spin-selective gap follows it and keeps the topological properties for the majority-spin states. This is the reason why the spin-selective topological phase is stabilized in a wide range of the chemical potential.

For example, as shown in the phase diagram (Fig. \ref{fig:phase}), we find the spin-selective topological phase for $1.2 \lesssim N_{tot} \lesssim 1.4$ at $U=24$. The corresponding changes in the electron number and the chemical potential are plotted in Figs. \ref{fig:ch} (b) and (c) at $U=24$. With increasing the electron number from $N_{tot}=1.2$, the chemical potential changes from $-2.91$ to $-1.96$ (Fig. \ref{fig:ch}(b)). This change is much larger than the gap size.
However, the above two effects cooperate and change the electron structure so that the change in electron number happens only through the down-spin (minority) electrons (Fig. \ref{fig:ch}(c)). As a result, the spin-selective gap remains as seen in Fig. \ref{fig:ch}(d). Thus we can confirm that the majority-spin states generate the gap, inducing non-trivial topological properties.

Systematically performing similar calculations, we finally arrive at the phase diagram already shown in Fig. \ref{fig:phase}; 
At half-filling, the system shows the antiferromagnetic order for $U>12$ and is the paramagnetic insulator for $U<12$. Note again that the paramagnetic Kondo insulator for $U<12$ is essentially the same as that found in refs. \cite{TKI_Dzero10,TKI_Dzero12,TKI_Tran12}. The remarkable point is that a topologically non-trivial spin-selective insulator is realized in a ferromagnetic metallic phase and is stabilized in a wide range of parameters.
\begin{figure}[!h]
\begin{center}
\includegraphics[width=80mm,clip]{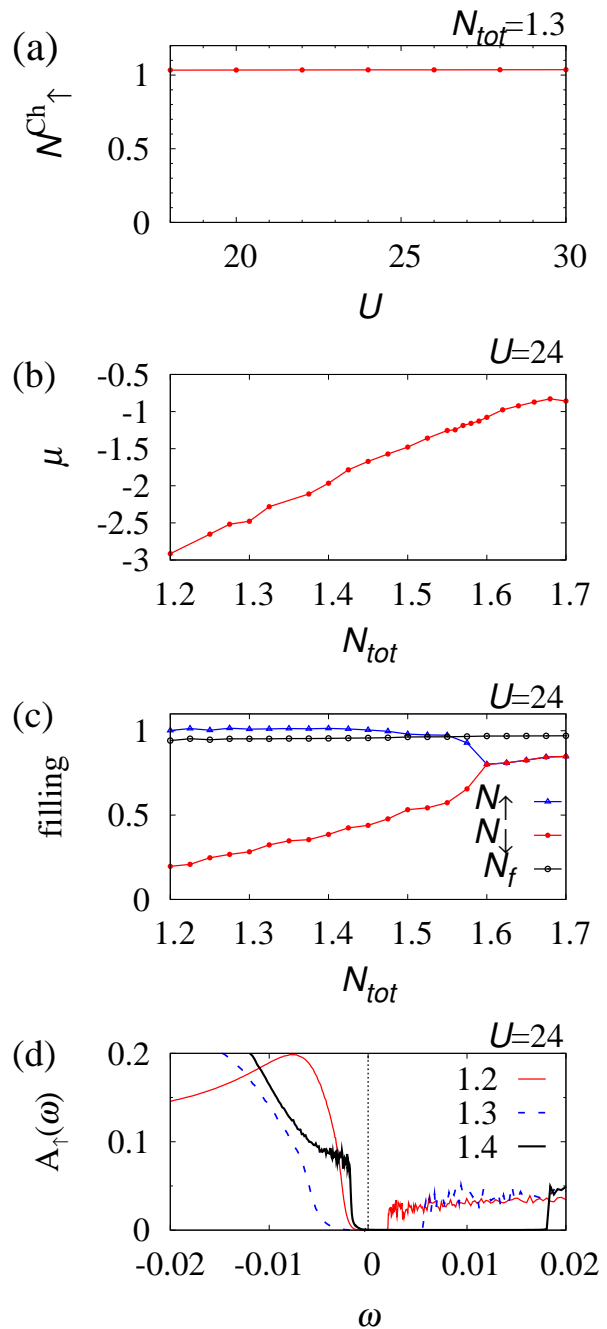}
\end{center}
\caption{
(a) Chern number for up-spin (majority) electrons. 
For $U<18$ and $U>30$, the topological invariant is not well-defined since the majority-spin state is metallic. (b) chemical potential as a function of the total filling (c) 
filling of the majority-spin states $N_{\uparrow}$ (the minority-spin state $N_{\downarrow}$, $f$-electrons $N_{f}$) as a function of the total filling $N_{tot}$ respectively. (d) Spectral functions for the majority-spin state $A_{\uparrow}=-\sum_{\alpha}\mathrm{Im}G^R_{\alpha,\alpha,\uparrow}/\pi$ around the Fermi energy $\omega=0$ (dotted line) for $N_{tot}=1.2,1.3$ and $1.4$, respectively.
}
\label{fig:ch}
\end{figure}

\section{Edge states}
Now, let us turn to the edge properties. 
In the spin-selective topological insulator found here, we expect the formation of gapless edge states resulting from non-trivial properties in the bulk. As clarified in refs. \cite{Gurarie11,Essin11}, the single-electron Green's function possesses  a pole or a zero at the open boundary, which separates the topological bulk and a trivial vacuum. The pole corresponds to the existence of a gapless edge, while the zero indicates the energy gap of the edge state. The non-zero Chern number in the bulk implies the existence of a chiral gapless edge (i.e. the Green's function does not have the zero at the boundary).  If the zero could emerge, the boundary state would be a Mott insulator, while the bulk is still a correlated band insulator. However, according the previous studies based on the DMFT, if the Mott transition occurs, the transition at the boundary occurs simultaneously with that at the bulk.\cite{TBI_Mott_Tada}
It is well known that such a chiral edge state in quantum Hall systems is described by the chiral-Luttinger liquid. 
However, an important difference is that the chiral edge state in our model is coupled to bulk ferromagnetic fluctuations, and this difference indeed results in a drastic change in the edge state.
In order to analyze the edge states, let us consider a semi-infinite system; we assume the spin-selective topological phase in $y>0$ and vacuum otherwise. Here, we phenomenologically introduce the following action for the edge state which is coupled to bulk ferromagnetic fluctuations via the exchange interaction:
\begin{eqnarray}
 S&=&S_{edge}+S_{c}+S_{mag} \label{eq: action}\\
 S_{edge}&=&-\frac{1}{4\pi}\sum_{n,k_x}\phi(i\omega_n,k_x) k_x(i\omega_n-vk_x) \phi(-i\omega_n,-k_x) \nonumber \\
 S_{c}&=&-g\sum_{n,{\bf k}}ik_x\phi(i\omega_n,k_x)\psi(-i\omega_n,-{\bf k})  \nonumber  \\
 S_{mag}&=&\sum_{n,{\bf k}}\psi(i\omega_n,{\bf k}){\chi'}^{-1}(i\omega_n,{\bf k})\psi(-i\omega_n,-{\bf k})  \nonumber  
\end{eqnarray}
\vspace{-3mm}
\begin{eqnarray}
 \chi'(i\omega_n,{\bf k})&=&\frac{1}{  \xi^{-2}+(k^2_x+k^2_y)+|\omega_n|/(\Gamma{\sqrt{k^2_x+k^2_y}})  },  \nonumber 
\end{eqnarray}
where $S_{edge}$, $S_{mag}$ and $S_{c}$ describe the Tomonaga-Luttinger liquid for the edge state, the bulk magnetic fluctuation and the coupling between them. $\phi(k)$ describes the boson field for the edge state, and the boson field $\psi$ describes the bulk spin fluctuations. 
$\xi$ and $\Gamma$ are the correlation length for the magnetic fluctuations and a constant parameter, respectively, and $g$ describes the exchange interaction.
By integrating out the $\psi$ field and using the relation $n_{k_x}\sim ik_x\phi(k_{x})$, we obtain the two-particle Green's function ($G(x,\tau)=-\langle T n(x,\tau)n(0)\rangle$). The retarded Green's function can be written as
\begin{eqnarray}
 G^{R}(\omega,k_x) = \frac{2\pi k^2_x} {k_x(\omega+i\delta-vk_x) +\pi g^2 k_x^2 \sum_{k_y} \chi'^R(k)}, 
\end{eqnarray}
where $(\chi'^R(\omega,{\bf k}))^{-1} = \xi^{-2}+(k^2_x+k^2_y)-i\omega/(\Gamma{\sqrt{k^2_x+k^2_y}})$. From this we can conclude that the edge state shows a non-Tomonaga-Luttinger liquid behavior. 
The dynamics shows a dissipative behavior because of the coupling with the bulk magnetic fluctuations. This implies that a composite collective magnetic mode, composed of the chiral Luttinger liquid and the spin fluctuation, emerges due to correlation effects. To characterize the anomalous dynamics, we here consider the spin-lattice relaxation time of the edge mode. Let us assume that a magnetic field is applied perpendicular to the $z$-axis.  Then, the longitudinal NMR relaxation time ($T_1$) is given as,
\begin{eqnarray}
 \frac{1}{T_1T}&=& C A^2 \lim_{\omega \rightarrow 0} \sum_{k_x} \frac{1}{\omega} \mathrm{Im} \chi^{zz}(k_x,\omega+i\delta) \nonumber \\
 &\sim& \frac{CA^2 2\pi^2g^2}{\xi^{-4}v'^2},
\end{eqnarray}
{
where $C$, $A$ and $v'$ denote a constant including the nuclear magnetic ratio, the hyperfine coupling constant, and the renormalized velocity, respectively. Here, we have used the relation, $\chi^{zz}(k_x,\omega+i\delta) \sim -G^{R}(k_x,\omega)$.  From these results, we find that the relaxation time decreases when the system approaches the ferromagnetic transition point ($T_{\mathrm{Curie}}=0$). Moreover, according to the self-consistent renormalization (SCR) analysis\cite{2D_Ferro_Hatatani95}, the contribution of the bulk ferromagnetic fluctuations in the two-dimensional system is $1/(T_1T) \sim \xi^{3}$, and the temperature dependence of correlation length is $\xi^{-2}\sim T^{2/3}e^{-4bT}$. Here, $b$ is constant which is proportional to square of the ferromagnetic moment at $T=0$. Therefore, if the magnetic fluctuations become stronger, the edge contribution becomes dominant and is experimentally observable; near the transition point ($T=0$), edge and bulk contributions are $1/(T_1T) \sim T^{-4/3}e^{8bT}$ and $1/(T_1T) \sim T^{-1}e^{6bT}$, respectively.
}

\section{Summary}
We have proposed a spin-selective topological insulator in a two-dimensional heavy-fermion system, which is realized in a ferromagnetic metallic phase. In this phase, a spin-selective gap is formed only in the majority-spin state, providing a topological insulating behavior. The many-body nature due to the RKKY interaction and the Kondo effect supports the robustness of the topological phase against a change in the chemical potential in a much larger range than the spin-selective gap. We have shown that the bulk topological structure gives the quantized Hall conductivity for the majority-spin electrons, thereby supporting the existence of a topologically protected chiral edge state. We have also discussed the effect of the bulk ferromagnetic fluctuations on the edge mode; the spin correlation function exhibits a non-Tomonaga-Luttinger liquid behavior with a large damping coming from bulk ferromagnetic fluctuations which may be detected by the NMR measurement. 
We expect that the two-dimensional heavy-fermion system studied in this paper can be realized in superlattice systems which have the inversion symmetry.\cite{Shishido_superlattice10}
Another possibility is doping topological Kondo insulators; so far, topological Kondo insulators are proposed for some heavy-fermion compounds\cite{TKI_Dzero10,TKI_Dzero12,TKI_Tran12}, which may be described by the half-filled Kondo lattice model. In these systems, we may have a chance to realize a spin-selective topological insulator even in a metallic phase.

\section{Acknowledgments}
This work is partly supported by JSPS through its FIRST Program, KAKENHI (Nos. 20102008, 23102714, 23540406), and the Global COE Program ``The Next Generation of Physics, Spun from Universality and Emergence'' from MEXT of Japan.
T. Y. thanks the Japan Society for the Promotion of Science (JSPS) for Research Fellowships for Young Scientists.

\end{document}